\documentstyle[prl,aps,epsf]{revtex}

\begin{document}
\title{
\draft
Divergent Signal-to-Noise Ratio and
Stochastic Resonance in Monostable Systems
}
\author{J. M. G. Vilar and J. M. Rub\'{\i}
}
\address{
Departament de F\'{\i}sica Fonamental, Facultat de
F\'{\i}sica, Universitat de Barcelona, \\
Diagonal 647, E-08028 Barcelona, Spain 
\date{\today}
}
\maketitle
\begin{abstract}
\leftskip 54.8pt
\rightskip 54.8pt
We present a class of systems for which the signal-to-noise ratio
always increases when increasing the noise and diverges at infinite
noise level.  This new phenomenon is a direct consequence of the
existence of a scaling law for the signal-to-noise ratio and implies
the appearance of stochastic resonance in some monostable systems.
We outline applications of our results to a wide variety of systems
pertaining to different scientific areas.  Two particular examples
are discussed in detail.
\end{abstract}
{\leftskip 54.8pt
}

Stochastic Resonance (SR)
\cite{Benzi,tri,Maki1,Maki2,libi,JSP,Moss,Wies,Wiese,thre,phi4} is a
phenomenon wherein the response of a system to a driven periodic
signal is enhanced at an optimized non-zero noise level.  Although,
increasing the noise level in order to enable us to more easily
detect a signal was considered counterintuitive, this constitutes one
of the most surprising results of the SR.  It seems obvious, however,
that the signal-to-noise ratio (SNR) must go to zero as noise is
increased indefinitely.  Contrarily, in this letter we present a
class of systems in which the SNR always increases when the noise is
increased and diverges at infinite noise level, instead of exhibiting
a maximum at a particular value of the noise.  This result implies
the presence of SR in monostable systems for which a maximum in the
SNR at non-zero noise level has never been observed before.  These
findings open up new possibilities concerning the application of SR
to a great variety of physical, chemical and biological systems.  To
be explicit we have applied our results to two particular cases;
namely, a ferromagnetic particle and a standard model of neural
excitable medium.

The class of systems we will discuss are described by only one
relevant degree of freedom whose dynamics is governed by the
following Langevin equation
\begin{equation}
\label{ourmodel}
{dx \over dt} = -h(t)x^{1+2n} + \xi(t) \;\; , 
\end{equation}
where $h(t)=k(1+\alpha\sin(\omega_0 t))$, with $k$ and $\alpha$
($<1$) constants, $n$ is an integer number and $\xi(t)$ is Gaussian
white noise with zero mean and second moment $\left< \xi(t) \xi(t+
\tau) \right> = D \delta(\tau)$, defining the noise level $D$.  The
system can be characterized by the quantity $v$ which is a function
of the variable $x(t)$.  This quantity is sometimes referred to as
the response to the oscillating force.  The effect of this force may
be analyzed by the power spectrum
\begin{equation}
P(\omega)=\int_0^{2\pi/\omega_0}dt \int_{-\infty}^\infty 
\left<v(t)v(t+\tau)\right>e^{-i\omega \tau}d\tau \;\; .
\end{equation}
To this purpose we will assume that it consists of a delta function
centered at the frequency $\omega_0$ plus a function $Q(\omega)$
which is smooth in the neighborhood of $\omega_0$ and is given by
$P(\omega) 
=Q(\omega)+S(\omega_0)\delta(\omega-\omega_0)$.  Then, the SNR is
defined by ${\rm SNR} \equiv {S(\omega_0)/Q(\omega_0)}$ and
consequently, has dimensions of inverse of time.

The existence of a characteristic time $\tau$ in our system will
enable us to propose the form of the SNR through the simple scaling
law
\begin{equation}
\label{SNRex}
{\rm SNR} = f(\alpha,\omega_0\tau)\tau^{-1} \;\; , 
\end{equation}
where $\tau^{-1}=D^{n/(1+n)}k^{1/(1+n)} $ and
$f(\alpha,\omega_0\tau)$ is a dimensionless function, provided that
$v(x)$ does not introduce another characteristic time.  We will
suppose that for a given value of $\tau$ the limit of SNR when
$\omega_0$ goes to zero exists.  As such, the following expression
for small driving frequencies holds
\begin{equation}
\label{SNRap}
{\rm SNR}\approx f(\alpha,0)\tau^{-1} . 
\end{equation}

Let us now discuss the main characteristics of our model upon varying
the exponent $n$.  If $n=0$, one finds the exact result ${\rm
SNR}=f(\alpha,\omega_0k^{-1})k$, which does not depend on the noise
level.  Even more interesting is the behavior obtained for the case
$n>0$.  The scaling of the SNR indicates that it increases when
increasing the noise level, achieving the behavior ${\rm SNR} \propto
D$ as $n$ goes to infinity.  A particular and common situation
illustrating this case corresponds to the potential $V_1(x)={1\over
4} h(t)x^4$ (Fig.  \ref{cuarticos}a), obtained when $n=1$, for which
SNR increases as $\sqrt D$.  In Fig.  \ref{cuarticos}b we have
depicted the SNR corresponding to $V_1(x)$.  Here the magnitude
giving the response of the system has been taken $v(x)=x^2$.  Our
result is obtained from numerical simulations by integrating the
corresponding Langevin equation by means of a standard second-order
Runge-Kutta method for stochastic differential equations
\cite{sde1,sde2}.  In order to verify the scaling law proposed
through eq.  (\ref{SNRap}) we have fit the values of the SNR to a
power law in the range of $D$ from $1$ to $1000$ for the potential
$V_1(x)$.  SNR is given by $aD^b$ with $b=0.498 \pm 0.004$ and
$a=0.76 \pm 0.02$, which is in good agreement with the theoretical
value $b={1\over2}$.  The power spectrum corresponding to $V_1(x)$
for two values of the noise levels is shown in Fig.
\ref{cuarticos}c.  Both the signal and noise background increase when
increasing $D$, but the signal increases faster.

In spite of its simplicity, our model encompasses a great variety of
common situations.  Around an equilibrium state most systems may be
approximated by a parabolic potential.  Thus, for $n=0$ our model
describes a system around an equilibrium state in a force field whose
intensity varies periodically in time.  A physical realization of
such a system could be a dipole under an oscillating field.  For
$n=0$, however, the SNR is independent of $D$.  In order to
understand the behavior of the SNR, as $D$ is increased we must take
into account the corrections to the parabolic approximation.
Commonly, these corrections are given by a term proportional to
$x^4$.  Two particular realizations of this situation have been
analyzed numerically resulting in an increase of the SNR (Fig.
\ref{cuarticos}b) with the noise level.  For
$V_2(x)=h(t)({1\over2}x^2+{1\over 4}x^4)$ (Fig.  \ref{cuarticos}a),
which basically corresponds to a potential that around the minimum
grows faster than a parabolic one, we expect that the SNR is an
increasing function of the noise, since for low noise level the
potential behaves as ${1\over2}h(t)x^2$, whereas for high noise level
as ${1\over4}h(t)x^4$.  A slightly different potential is
$V_3(x)={1\over 2}x^2+{1\over 4} h(t)x^4$ (Fig.  \ref{cuarticos}a),
which only differs from the previous one in the behavior at low noise
level.  Since under this circumstance this potential reduces to
${1\over2}x^2$ and consequently it is not modulated by $h(t)$, SNR
goes to zero for low $D$.  For $n=1$, our model describes the
dynamics of a system at the critical point of both the pitchfork and
Hopf bifurcations\cite{chus}, occurring in many systems including, to
mention just a few\cite{Haken,Murray}, chemical reactions, models of
populations, convection in liquids, lasers and instabilities in
semiconductors.

The divergence of the SNR is due to the fact that the potential is
unbounded.  It is obvious that for a bounded potential the noise can
completely destroy the response of the system.  An important
consequence follows from the previous results:  if the SNR grows for
low noise level (when the potential around the minimum can be
approximated by a potential like $V_2$ or $V_3$) and the potential is
bounded (which implies that SNR goes to zero for large noise) then
the SNR exhibits a maximum, thus indicating the appearance of SR.  To
illustrate this point we report results of numerical simulations for
the dynamics of a ferromagnetic particle\cite{dau,Agus} under an
external magnetic field and with energy of anisotropy
$h(t)sin^4\theta$, with $\theta$ being the angle between the magnetic
moment and the axis of easy magnetization.  The external magnetic
field is then applied in the direction of the easy axis of
magnetization and its intensity is as high as the system becomes
monostable.  The dynamics of the magnetization may be described by
the Langevin equation
\begin{equation}
{d \theta \over dt} = -\sin\theta 
-h(t)\sin^3\theta\cos\theta +\xi(t) \;\; ,
\end{equation}
where the first term on the right hand side accounts for the
interaction with an external magnetic field, the second for the
anisotropic effects and the third is a noise source due to a random
field or to thermal fluctuations.  The parameter $h(t)$ is assumed to
be oscillatory, the reason being, for example, that the presence of
oscillations of the pressure of the medium surrounding the particle.
The response of the system is now given by $\cos\theta$, i.e.  by the
magnetization.  For low noise, the potential (Fig.  \ref{ferro}a) and
the magnetization reduces to
${1\over2}\theta^2+{1\over4}h(t)\theta^4$ and $1-{1\over2}\theta^2$,
respectively.  Therefore our previous results apply to this case.
The corresponding SNR is shown in Fig.  \ref{ferro}b and exhibits a
maximum at a finite noise level.  This result clearly shows the
existence of an optimal noise level for which the system is more
sensible to periodic changes of the environment.  In Fig.
\ref{ferro}c we have represented the power spectrum for some values
of the noise level.

The next example to be considered corresponds to a standard model of
a neural excitable medium.  This model characterizes the activity
generated in a slab of neural tissue comprising a very large number
of closely packed and coupled nerve cells\cite{Cow1,Cow2,Cow3}.  We
will consider the case of all-to-all connectivity in which spatial
dependence may be ignored The model of neural excitable medium is
given by the following equation\cite{Cow1}
\begin{equation}
C{dU \over dt}=-R^{-1}U+\phi(U)+P 
 \;\; ,
\end{equation}
describing the dynamics of the spatial average of the transmembrane
potential $U$.  Here $C$ is the membrane capacitance, $R$ the
membrane resistance and $P$ is an external current applied to the
net.  The nonlinear term $\phi(U)$ is proportional to the gain of the
neuron and accounts for its mean firing rate.  Its form is usually
taken to be a sigmoidal, e.g.
\begin{equation}
\phi(U)=\varepsilon(1+e^{-\nu(U-\theta)})^{-1} \;\;,
\end{equation}
where $\nu$ is a constant, fixing the sensitivity to excitation of
the population, $\theta$ is the threshold mean voltage and
$\varepsilon$ a parameter depending on the structure of the net and
the characteristics of the neuron.  We will consider that the
external current applied to the net fluctuates and that it may be
approximated by a Gaussian white noise ($\left<P(t)\right>=0$ and
$\left<P(t) P(t+ \tau) \right> = D \delta(\tau)$) The potential
function (Fig.  \ref{neuron}a) corresponding to the variable $U$ is
given by
\begin{equation}
V(U)={1 \over C}\left({1\over2}R^{-1}U^2
-{\varepsilon \over \nu }
\ln\left(1+e^{\nu\left(U-\theta\right)}\right)\right) \;\; .
\end{equation}
The membrane resistance $R$ can be modified through small changes in
the permeability of a suitable ion.  Let us consider that this
modification is periodic in time.  There exists a range of parameters
for which the neurons of the net are not excited, i.e.  $U=0$,
consequently this state is not affected by small variations of $R$.
The situation changes drastically with the addition of noise.  For
small values of the noise, when the term $\phi(U)$ is irrelevant, the
dynamics of this model may be described by eq.  (\ref{ourmodel}),
with $n=0$.  In this case, the SNR for $v=U^2$ is independent of $D$,
but in this system the interesting variable is $U$ instead of $U^2$.
For small noise values, the potential is symmetric in $U$, thus
giving a zero SNR for the variable $U$.  When increasing the noise
level the symmetry is lost, since positive fluctuations of $U$ may
cause firing of the neurons.  Variations of $R$ then modulates the
amplitude of the fluctuations of $U$, giving rise to periodicity in
the firing rate of the neurons, which implies a non-vanishing SNR.
Further increasing of the noise level then leads to the restoration
of the symmetry.  Consequently, this model exhibits SR as has been
shown in Fig.  \ref{neuron}b.  It is interesting to remark that, in
contrast with previous results concerning the appearance of SR for a
single neuron\cite{neu1,neu2}, our results refer to the prediction of
the phenomenon for an ensemble of neurons.

Our analysis has revealed the presence of SR in a wide variety of
situations, embracing different scientific areas, which have not been
considered up to now.  A methodological aspect to be emphasized, and
that could be used in subsequent studies, is that arguments as simple
as scaling laws or considerations about the symmetry may help us in
predicting the enhancement of signals via SR.  Our work, then, offers
new perspectives on what concerns the consideration of SR as a
general phenomenon that might apply to diverse systems.

The authors would like to thank A.  P\'erez-Madrid and Ricard V.
Sol\'e for fruitful discussions.  This work was supported by DGICYT
of the Spanish Government under Grant No.  PB92-0859.  One of us
(JMGV) wishes to thank the Generalitat de Catalunya for financial
support.

\begin{figure}[th]
\caption[a]{\label{cuarticos}
{\bf (a)} Potentials $V_1(x)$, $V_2(x)$ and $V_3(x)$, for the maximum
value of $h(t)$ (solid line) and for the minimum value (dashed line).
Here $k=1$, $\alpha=0.5$ and $\omega_0/2\pi=0.1$.
{\bf (b)} Behavior of the SNR for the three potentials presented
previously.
{\bf (c)} Power spectrum corresponding to $V_1(x)$ for $D=0.1$ and
$D=1000$.}
\end{figure}

\begin{figure}[th]
\caption[b]{\label{ferro}
{\bf (a)} Potential energy of the ferromagnetic particle (eq.  (5))
as function of $\theta$ for the maximum value of $h(t)$ (solid line)
and for the minimum (dashed line).  The parameters are taken $k=0.3$,
$\alpha=3/2$ and $\omega_0/2\pi=0.1$.
{\bf (b)} ${\rm SNR}$ for the previous values of the parameters
obtained through computer simulations.
{\bf (c)} Power spectrum for $D=0.05$, $0.15$, $0.35$ and $1$.}
\end{figure}

\begin{figure}[th]
\caption[c]{\label{neuron}
{\bf (a)} Representation of the potential function $V(U)$ for the
maximum value of $R^{-1}=h(t)$ (solid line) and for the minimum value
(dashed line).  The parameters used here are $C=1$, $\theta=2$,
$\varepsilon=1$, $\nu=10$ and the resistance $R^{-1}=h(t)$, with
$k=2$, $\alpha=0.5$ and $\omega_0/2\pi=0.01$
{\bf (b)} SNR for the previous parameters.}
\end{figure}


\begin{references}

\bibitem{Benzi} R. Benzi,  A.  Sutera, and A. Vulpiani,
J. Phys. A {\bf 14}, L453-L457 (1981).

\bibitem{tri} S. Fauve and F. Heslot,
Phys. Lett. A {\bf 97} 5-7 (1983).

\bibitem{Maki1}B. McNamara, K. Wiesenfeld, and R. Roy,
Phys. Rev. Lett. {\bf 60}, 2626-2629 (1988).

\bibitem{Maki2}B. McNamara and K. Wiesenfeld,
{\it Phys. Rev. A} {\bf 39}, 4854-4869 (1989).

\bibitem{libi}A. Simon and A. Libchaber,
Phys. Rev. Lett. {\bf 68} 3375-3378 (1992).

\bibitem{JSP}Proceedings of the NATO Advanced Research
Workshop on Stochastic Resonance, San Diego,
1992 [J. Stat. Phys. {\bf 70}, 1 (1993)]. 

\bibitem{Moss}F. Moss, in
{\it Some Problems in Statistical Physics},
edited by G. H. Weiss (SIAM, Philadelphia,1994).

\bibitem{Wies}K. Wiesenfeld,  D. Pierson, E. Pantazelou, 
C. Dames,  and F. Moss,
Phys. Rev. Lett. {\bf 72}, 2125-2129 (1994).

\bibitem{Wiese} K. Wiesenfeld and  F. Moss,
Nature {\bf 373}, 33 (1995).

\bibitem{thre}Z. Gingl, L. B. Kiss, and F. Moss,
Europhys. Lett. {\bf 29} 191-196 (1995).

\bibitem{phi4}F. Marchesoni, L. Gammaitoni, and A. R. Bulsara,
Phys. Rev. Lett. {\bf 76} 2609-2612 (1996).

\bibitem{sde1}P. E. Kloeden and R. A. Pearson,
J. Austral. Math. Soc., Ser B, {\bf 20} 8-12 (1977).

\bibitem{sde2}J. R. Klauder and W. P. Petersen,
SIAM J. Numer. Anal. {\bf 22} 1153-1166 (1985).

\bibitem{chus}H. Schuster,
{\it Determinisc Chaos} (VCH, Weinheim, 1989).

\bibitem{Haken}H. Haken,
{\it Synergetics} (Springer-Verlag, New York, 1983).

\bibitem{Murray}J. D. Murray, {\it Mathematical Biology},
(Springer-Verlag, New York, 1989).

\bibitem{dau}L. D. Landau and E. M. Lifshitz,
{\it Electrodynamics of Continuous Media}
(Pergamon, New York, 1981), Vol 8.

\bibitem{Agus}A. Per\'ez-Madrid and J.M. Rub\'{\i},
Phys. Rev. E {\bf 51} 4159-4164 (1995).

\bibitem{Cow1}J. D. Cowan and G. B. Ementrout,
in {\it Studies in Mathematical Biology, Part I},
edited by S. A. Levin
(The Mathematical Association of America, Washington, 1978).

\bibitem{Cow2}H. R. Wilson and J. D. Cowan, 
Biophys. J. {\bf 12} 1-24 (1972).

\bibitem{Cow3} J. L. Fedelman and J. D.  Cowan,
Biol. Cybern. {\bf 17}, 39-51 (1975).

\bibitem{neu1} A. Longtin, A. Bulsara, and F. Moss,
Phys. Rev. Lett. {\bf 67} 656-659 (1991).

\bibitem{neu2}J. K. Douglass, L. Wilkens, E. Pantazelou, and F. Moss,           
Nature {\bf 365} 337-340 (1993).


\end{references}
\end{document}